\documentclass[aps,pre,twocolumn,floats,floatfix,english]{revtex4}
\usepackage[T1]{fontenc}
\usepackage[latin1]{inputenc}
\usepackage{amsmath}
\usepackage{graphicx}
\usepackage{setspace}
\usepackage{epsfig}
\makeatletter

\providecommand{\LyX}{L\kern-.1667em\lower.25em\hbox{Y}\kern-.125emX\@}

\usepackage{babel}
\makeatother
\begin{document}
\title{Biology helps you to win a game}

\author{Marko Sysi-Aho}
\author{Anirban Chakraborti}
\email{anirban@lce.hut.fi}
\homepage{http://www.lce.hut.fi/~anirban}
\author{Kimmo Kaski}

\affiliation{Laboratory of Computational Engineering, Helsinki
University of Technology, \\
P. O. Box 9203, FIN-02015 HUT, Finland.}

\begin{abstract}

We present a game of interacting agents which mimics the complex dynamics found in many natural and social 
systems. These agents modify their strategies periodically, depending on their 
performances using genetic crossover mechanisms, inspired by biology. We study 
the performances of the agents under different conditions, and how they 
adapt themselves. In addition the dynamics of the game is investigated.
 
\end{abstract}

\maketitle
\section{Introduction}
Is the ``survival of the fittest'' principle limited to biology only? Perhaps 
not and there could be other spheres of life in which this principle is 
applicable. Competition plays a key role and in order to compete and thus survive in any 
environment or situation, one primarily needs to adapt in order to succeed. 
Then what is adaptation and evolution? Adaptation is an alteration or 
adjustment in 
structure or habits, often hereditary, by which a species or individual improves
its condition in relationship to its environment. Evolution is the change 
in the genetic composition of a population during successive generations, as a 
result of natural selection acting on the genetic variation among individuals, 
and resulting in the development of a new species.
Here, we show that in the behaviour of various complex systems found in 
natural and social environments \cite{parisi,huberman,nowak,lux,arthur}, that 
can be characterized by 
the competition among interacting agents for scarce resources,
adaptation to the environment plays a very important role.

These agents could be
diverse in form and in capability, ranging for example, from carcinogenic cells in the
human body to multinational firms in the global financial market. In these dynamically
evolving complex systems the nature of agents and their behaviour
differ a lot but they have a common underlying mechanism. In order to have a deeper 
understanding of the interactions of the large number of agents, one  
should first consider the individual capabilities of the agents. Its behaviour may 
be thought of as a collection of simple rules governing ``responses'' to 
numerous ``stimuli''.
The rules of action serve as the agents' strategies, and the
behaviour of an agent is the rules acting sequentially.
Therefore, in order to model any complex dynamically adaptive system, a major concern is the
selection and representation of the stimuli and responses, since the behaviour
and strategies of the component agents are determined thereby. 
Then the agent needs to adapt to different situations, where the 
experience of an agent guides it to change its structure so that as time 
passes, the agent learns to make better use of the environment for its own 
benefit.
However, the timescales over which the agents adapt vary from one individual to 
another and also from one system to another.

In complex adaptive systems, many interesting temporal patterns are produced, 
since a major part of the environment of a particular agent includes other
adaptive agents and a considerable amount of agent's effort goes
in adaptation and reaction to the other agents. 
Thus the situation is considerably different and more complicated than in
game theory \cite{game} and conventional theories in economics, where the study is of patterns in behavioural equilibrium that 
induce no further interaction.

In this paper, we study a simple game based on the basic minority game
\cite{challet1,challet2,cavagna,riolo,lamper}, where the agents adapt 
themselves by modifying their strategies from time to time, depending on their 
current performances, using genetic crossover mechanisms 
\cite{holland,goldberg,lawrence,Marko1}. The game can be a very simple 
representation of a complex adaptive system. We make a comparative study of 
their performances with the various mechanisms and in a ``test'' situation. 

\section{Model}
In this section we give a brief description of the model. The basic minority
game  consists of an odd number $N$ of agents who can perform at a given time 
$t$, any of the two possible actions denoted here by $0$ or $1$. The 
minority game was based on the El Farol bar problem, created by Brian Arthur, 
in which a population of agents have to decide whether to go to the bar every
Thursday night, and so there were two possible actions ``to attend'' denoted by
$1$ and ``not to attend'' denoted by $0$, depending on whether the bar was too 
crowded or not \cite{arthur}. 
An agent wins the game if it is one of the members of the minority
group. 
All the agents are assumed to have access to finite amount of {}``global''
information: a common bit-string {}``memory'' of the $M$ most recent
outcomes. With this there are  $2^M$ possible ``history''
bit-strings. Now, a {}``strategy'' consists of two possible
responses, which in the binary sense are an action $0$ or 
action $1$ to
each possible history bit-strings. Thus, there are $2^{2^{M}}$
possible strategies constituting the whole {}``strategy space''.

\begin{figure}
\epsfig{file=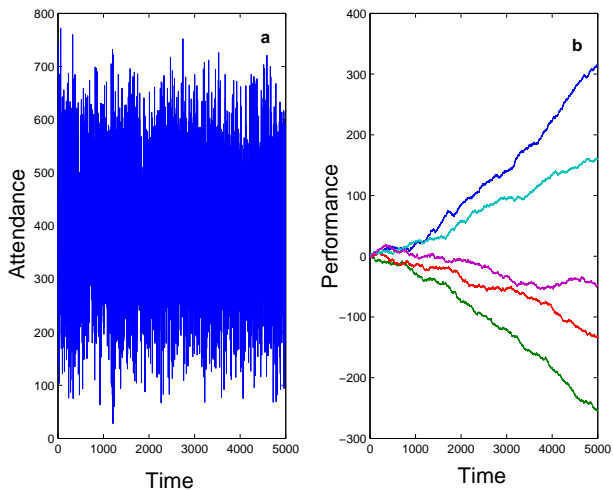,width=3.2in}
\caption{
Plots of (a) attendance and (b) performance of the players for the basic 
minority game with $N=801$, $M=6$, $k=10$ and  $T=5000$.
}
\label{bmg}
\end{figure}

Each time
the game has been played, time $t$ is incremented by unity and one {}``virtual'' point is assigned to the strategies that 
predicted the correct outcome and the best strategy is the one which has
the highest virtual point score. The performance of a player is measured
by the number of times the player wins, and the strategy, which the
player uses to win, gets a {}``real'' point. The number of agents
who have chosen a particular action, say $1$ which represents ``to attend'', 
is denoted by
$A_{1}(t)$ (also referred as ``attendance'') and it varies with time. We have plotted the attendance and performance 
for the basic minority game in Fig. \ref{bmg}.

Now we define the total utility of the system as the number of persons in the 
minority group at a given time $t$. For convenience, we mathematically define a scaled utility (total utility/maximum utility) as

\begin{equation}
U=[(1-\theta(x_t-x_M))x_t+\theta(x_t-x_M)(N-x_t)]/x_M,
\end{equation}

\noindent where $x_M=(N-1)/2$, 
$x_t$ is either equal to $A_1(t)$ or $A_0(t)$, and $\Theta (x)$ is Heaviside step function:

\begin{displaymath}
\theta(x)=\left\{ \begin{array}{ll}
              0 & \textrm{ when $x \le 0$} \\
             1 & \textrm{ when $x > 0$}.
\end{array}\right.
\end{displaymath}

\begin{figure}
\epsfig{file=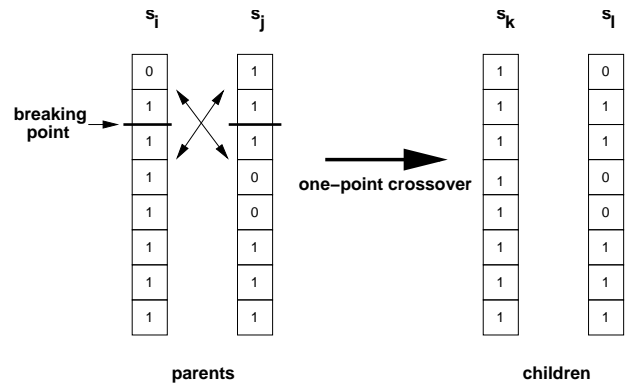,width=3.2in}
\caption{ 
Schematic diagram to illustrate the mechanism of one-point genetic
crossover for producing new strategies. The strategies $s_{i}$ and
$s_{j}$ are the parents. We choose the breaking point randomly and
through this one-point genetic crossover, the children $s_{k}$ and
$s_{l}$ are produced.
}
\label{cross}
\end{figure}

The players examine their performances after every time interval $\tau $. 
If a player finds that he is among
the fraction $n$ (where $0<n<1$) who are the worst performing
players, he adapts himself and modifies his strategies. The mechanism by which
the player creates new strategies is genetic crossover, whereby 
he selects the two {}``parents'' from his pool of $k$ strategies and
creates two new {}``children'' \cite{lawrence, Marko1}, as described in 
Fig. \ref{cross}.

If the parents are chosen randomly from the pool of strategies then the 
mechanism represents a ``one-point genetic crossover'' and if the parents are the 
best strategies then the mechanism represents a ``hybridized genetic crossover''.
The children may replace parents or two worst strategies and accordingly four different 
interesting cases arise:
(a) one-point genetic crossover with parents ``killed'', i.e. parents are replaced by the children,
(b) one-point genetic crossover with parents ``saved'', i.e. the two worst 
strategies are replaced by the children but the parents are retained,
(c) hybridized genetic crossover with parents ``killed''
 and (d) hybridized genetic crossover with parents ``saved''.

It should be noted that the mechanism of evolution
of strategies is considerably different from earlier attempts \cite{challet1,li1,li2}. This is because in this mechanism the strategies are changed by the agents 
themselves and even though the strategy space evolves continuously, its size
and dimensionality remain the same.

The Hamming distance $d_H$ between two bit-strings is defined as the ratio of 
the number of uncommon bits to the total length of the bit strings. It is a 
measure of the correlation between two strategies: 

\begin{displaymath}
d_H=\left\{ \begin{array}{lll}
              0 & \textrm{ correlated} \\
             0.5 & \textrm{ uncorrelated}\\
             1 & \textrm{ anti-correlated}
\end{array}\right.
\end{displaymath}

\noindent
which can be plotted as the game evolves.

\section{Results}

\begin{figure}
\epsfig{file=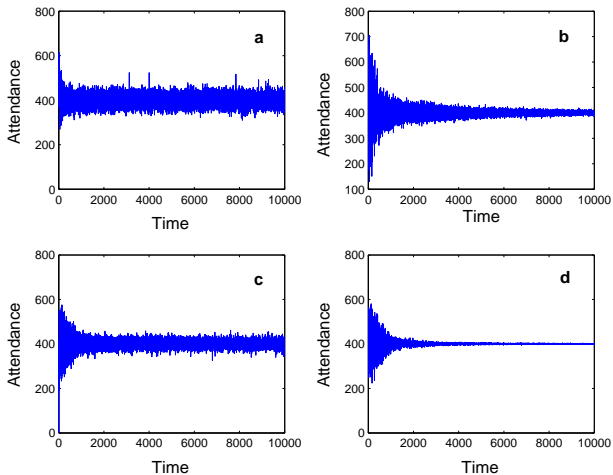,width=3.2in}
\caption{ Plots of the attendances by choosing parents
randomly (a) and (b), and using the best parents in a player's pool
(c) and (d). In (a) and (c) case parents are replaced by children and
in (b) and (d) case children replace the two worst strategies. 
Simulations have been done with 
$N=801$, $M=6$, $k=16$, $t=40$, $n=0.4$ and  $T=10000$.} 
\label{all4a}
\end{figure}

In order to determine
which mechanism is the most efficient, we have made a comparative study of the 
four cases, mentioned earlier. We plot the attendance as a function of time for
the different mechanisms in Fig. \ref{all4a}.

\begin{figure}
\epsfig{file=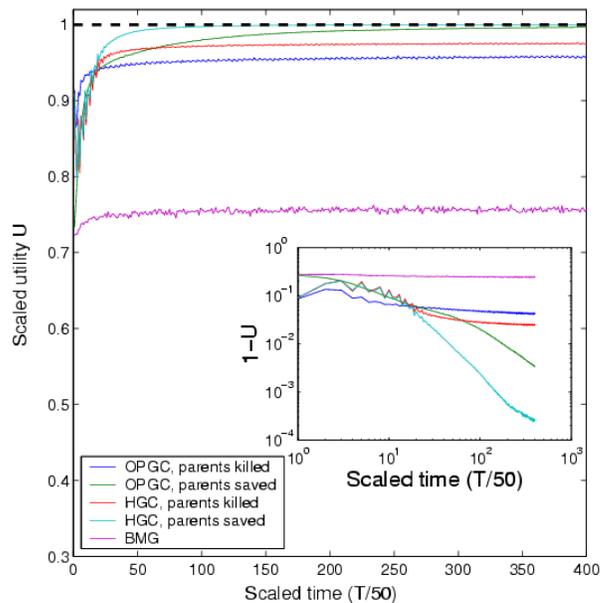,width=3.2in}
\caption{ Plots of the scaled utilities of the four different mechanisms in comparison with that of the basic minority game.
Each curve represents an ensemble average over 100 runs
and each point in a curve is a time average over a bin
of length 50 time-steps. In the inset, the quantity ($1-U$) is plotted against  scaled time in the double logarithmic scale. Simulations are done with $N=801$, $M=6$, $k=16$, 
$t=40$, $n=0.4$ and  $T=20000$.
 \label{all4b}}
\end{figure}

In Fig. \ref{all4b} we show the total utility of the system in each of the cases (a)-(d), where we
have plotted results of the average over 100 runs and each point in the utility
curve represents a time average taken over a bin of length 50 
time-steps. The simulation time is 
doubled from those in Fig. \ref{all4a}, in order to expose the asymptotic 
behaviour better. 
On the basis of Figs. \ref{all4a} and \ref{all4b}, we find that the case (d)
is the most efficient.

\begin{figure}
\epsfig{file=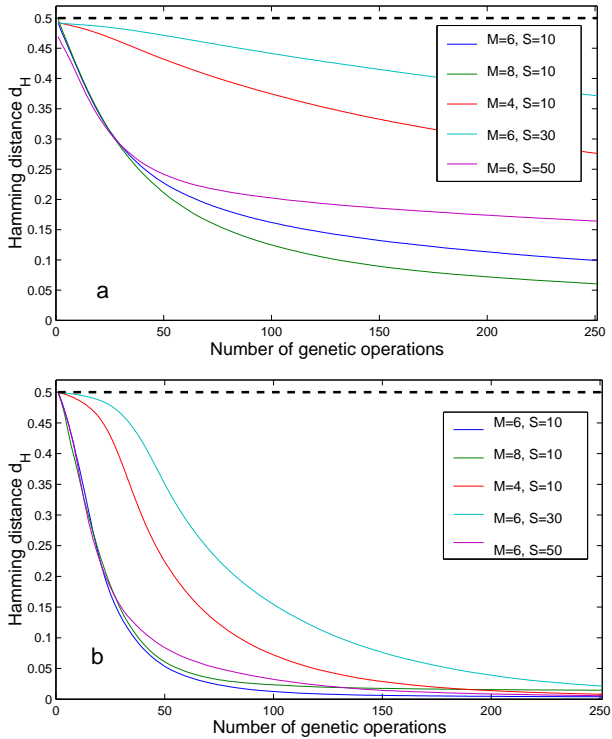,width=3.2in}
\caption{
Plot of the average Hamming distance of all the strategies in a pool of a player
with time, where the player adapts using (a) one-point genetic crossover and (b)
hybridized genetic crossover, and in both cases the two
worst strategies are replaced by the children and the parents are also saved.
Each curve is an ensemble average over 20 runs.}
 \label{ham1}
\end{figure}

In Fig. \ref{ham1} (a) one can see the evolution of the average Hamming distance
of all the strategies of a player in a game, where the player adapts using 
one-point genetic crossover and the two worst strategies are replaced by the children and the parents are also saved. 
It should be noted that the Hamming distance
can change only when the worst strategies 
are replaced by the children and the parents are saved, where the bits in a 
strategy pool can change over time. Otherwise the bits in the pool of strategies
remain the same. 
We observe that the curves tend to move downwards from around $0.5$ towards zero, which means that as the time
evolves, the correlation amongst the strategies increases and the strategies in the pool of a particular agent converges
towards one strategy. The nature of the curves depend a lot on the parameters of the game.
In Fig. \ref{ham1} (b) one can see the evolution of the average Hamming distance
of all the strategies of a player in the game, where the player adapts using 
hybridized genetic crossover and the two worst strategies are replaced by the 
children and the parents are also saved. Here too, the strategies in the pool of a
particular agent converges towards one strategy, and at a faster rate than with the 
previous mechanism.
We observe that increasing memory $M$ does not change dramatically
the convergence rate, but as we increase the number of strategies in
the pools, the convergence slows down.

\begin{figure}
\epsfig{file=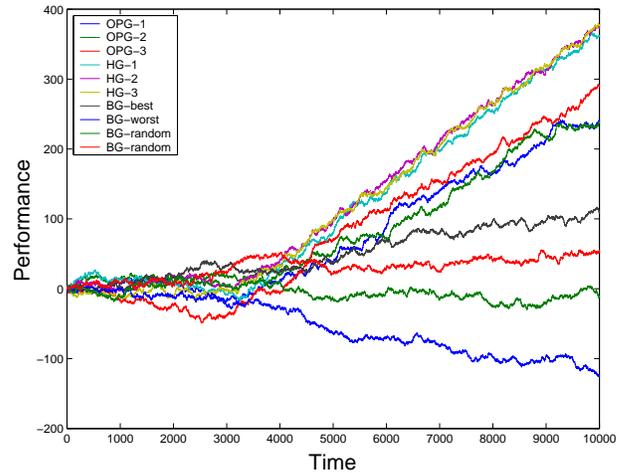,width=3.2in}
\caption{
Plot of the performance of the players where after $T=3120$ time-steps, six players begin 
to adapt and modify their strategies: three using hybridized genetic crossover 
mechanism and the other three using one point genetic crossover, where children 
replace the parents.
Other players play the basic minority game all the time and do not adapt. 
The simulations are done with $N=801$, $M=8$,
$k=16$, $n=0.3$, $t=80$, and $T=10000$.}
 \label{mixed2}
\end{figure}

In order to investigate what happens in the level of an individual agent, we
created a competitive surrounding-- ``test'' situation where
after $T=3120$ time-steps, six players begin
to adapt and modify their strategies such that three are using hybridized genetic crossover
mechanism and the other three one point genetic crossover, where children
replace the parents.
The rest of the players play the basic minority game. In this case it turns out that
in the end the best players are those who use the hybridized mechanism,
second best are those using the one-point mechanism, and the bad
players those who do not adapt at all.
In addition it turns out that the competition amongst the players who adapt using the hybridized genetic
crossover mechanism is severe.

\section{Conclusion}

We can summarize our findings by stating that adaptation improves not only the individual player's 
performance but also improves the total utility of the system. The best results
are found for the players who adapt and modify their strategies using the 
hybridized genetic crossover mechanism and the children replace the two worst
strategies and the parents are saved. The mechanism of adaptation is very simple
and can be used to model different complex adaptive systems. It can also be 
potentially developed to include other features like mutation. We can thus say 
that in a way, ``biology helps you to win a game''.

\begin{acknowledgments}
This research was partially supported by the Academy of
Finland, Research Centre for Computational Science and Engineering,
project no. 44897 (Finnish Centre of Excellence Programme 2000-2005).
\end{acknowledgments}

\end{document}